# High-quality NiFe thin films on oxide/non-oxide platforms via pulsed laser deposition at room temperature

H. Yan, G. J. Omar[*], Z. T. Zhao, Lim Zhi Shiuh, A. Ariando[*]

*Department of Physics, Faculty of Science, National University of Singapore, Singapore 117551, Singapore*

*Email: ariando@nus.edu.sg

**Abstract**

Soft ferromagnetic NiFe thin films are promising for applications in spintronic devices because of their constituent electrical and magnetic properties. Electron beam evaporation and sputtering techniques have been used to deposit NiFe thin films. For in-situ stacking of NiFe with functional complex oxides, the pulsed laser deposition (PLD) method is highly desirable. However, the growth of high-quality NiFe (and non-oxide thin films in general) by PLD remains a formidable task. Here, we report high-quality NiFe thin films of various thicknesses on oxide/non-oxide substrates with desirable magnetic properties by PLD at room temperature. The magnetic properties are found to be strongly dependent on the laser fluence of the deposition process. The laser fluence of 4 Joule/cm$^2$ produces the highest magnetization of ~547 emu/cc. The small coercivity (few Oersted) and sharp ferromagnetic switching behaviour indicate uniaxial anisotropy with an easy axis along the in-plane direction. In addition, thickness-dependent magnetodynamics characterizations are studied via ferromagnetic resonance. Our findings indicate the ferromagnetic characteristics are sensitive to the quality of the oxide/non-oxide substrate surface. These results offer significant insight into the PLD-based development of thin metal magnetic films.

**Keywords:** Pulsed laser deposition, ferromagnetic resonance, magnetization dynamics, NiFe films

**Introduction**

Since Arnold and Elmen originally discovered permalloy (NiFe) in 1923, it has attracted tremendous research interest because of its highly desirable properties, such as high permeability, low coercivity, and near-zero magnetostriction.[1] After decades, thin films of permalloy were reported in the 1950s.[2-4] The essential features of thin permalloy films that anisotropy fields of a few Oe and significant saturation magnetization were obtained.[4] Due to the attractive properties of thin NiFe films, it has been used as a constituent of many thin-film devices, such as free layers of giant magnetoresistance,[5] spin injection and accumulation,[6] and ferromagnetic resonance experiment[7] in spintronics devices.

Previously, thin NiFe films were always deposited by electron beam evaporation[8] and sputtering techniques.[9-11] The anisotropy field can be controlled in various ways, such as deposition in a moderate magnetic field (~100 Oe),[12] substrates,[10] and oblique deposition.[13, 14] In the last decades, pulsed laser deposition (PLD) has emerged as one of the most popular and intrinsically simple techniques for depositing a wide range of exciting functional materials. The PLD technique is often used to deposit multi-component oxide films.[15, 16] Depositions of NiFe by PLD, however, are surprisingly sparsely reported in the literature. Randolph *et al*. described the PLD growth of NiFe/Ag superlattices.[17] However, they showed that the NiFe films change from ferromagnetic to non-ferromagnetic at thicknesses less than ~20 nm.

In this work, we demonstrate the high-quality growth of soft ferromagnetic NiFe thin films on various oxide/non-oxide substrates by PLD at room temperature. Using a stoichiometric target with the chemical formula $Ni_{80}Fe_{20}$, thin films were deposited in high vacuum at room temperature.

Merits of this approach for growing oxide and non-oxide thin films are discussed. Atomic Force Microscopy (AFM) investigations show clean and flat surfaces. The static magnetic measurement indicates the ferromagnetic nature of the thin films. In addition, ferromagnetic resonance (FMR) measurement shows the dynamic properties and determines the damping parameters with respect to the thickness of the thin films and substrates.

**Experimental details**

High-quality NiFe thin films were grown by PLD at room temperature and at a high vacuum base pressure of ~$6\times10^{-8}$ mbar to avoid any *in-situ* oxide formation during the deposition. The NiFe target was ablated by a KrF excimer laser beam (248 nm, 30 ns FWHM) operating at a repetition rate of 5 Hz. The laser fluence ($F_L$) for each deposition was controlled from low power of 1.0 J/cm$^2$ to a maximum possible high power of 4.0 J/cm$^2$. The NiFe thin films with various thicknesses ($t$) were deposited at pressures <$4\times10^{-7}$ mbar (high Vacuum) at room temperature on various oxide/non-oxide substrates, namely, HF-treated-SrTiO$_3$ (HF-STO) (001), KTaO$_3$ (KTO) (001), pristine-SrTiO$_3$ (001) (STO), LaAlO$_3$ (LAO) (001), silicon (001) and Quartz. The surface roughness of the former four kinds of oxide substrates is all around 120-300 pm, whereas it is about 800-900 pm for the two kinds of non-oxide substrates. The film thickness was controlled by the number of laser pulses during deposition. The substrates were cleaned with acetone and attached to the holder 7 cm away from the target inside the main vacuum chamber. A 2 nm Al$_2$O$_3$ oxide capping layer was grown *in-situ* to prevent *ex-situ* oxidation of the NiFe films. The surface morphology of the films was characterized by Atomic Force Microscopy (AFM). Magnetic characterization was carried out using Magnetic Property Measuring System (MPMS) System fitted with a Superconducting Quantum Interference Device (SQUID) attachment by applying a magnetic field along both in-plane (IP) and out-of-plane (OOP) directions at room temperature. The ferromagnetic resonance (FMR) measurements were carried out by NanOsc Instruments Cryo ferromagnetic resonance (FMR) in Quantum Design Physical Property Measurement System (PPMS) at 300 K. Frequencies of 2-40 GHz were swept while applying different ranges of magnetic fields.

**Results and discussion**

Figure 1(a) shows the degree of unevenness of the film surface (morphology) in a scan area of 2×2 µm$^2$ of the NiFe films grown on silicon at $F_L$=4.0 J/cm$^2$ with $t$=12 nm. The roughness of the film surface is ~0.263 nm. Figure 1(b) shows the room-temperature resistivity $\rho$ of the obtained structures as a function of the $F_L$ used for the NiFe growth. For samples grown at low $F_L$ (< 3 J/cm$^2$), the resistivity $\rho$ is beyond the measurement limit of our PPMS transport measurement tool. Whereas the thin films are observed to be conducting above a certain high $F_L$ fluence threshold (⩾3 J/cm$^2$). Laser fluence ($F_L$) plays a critical role in the growth of high-quality NiFe films. The resistivity $\rho$ of thin films on silicon as a function of thickness for $F_L$ = 3 and 4 J/cm$^2$ is shown in Fig. 1(c). It is noteworthy that below 10 nm of film's thickness ($t$), the resistivity increases by nearly two orders of magnitude while the behaviour of the resistivity remains almost constant (around 10 µΩcm) for thicker films. In addition, the resistivity $\rho$ of NiFe is inversely related to the Fluence $F_L$. This suggests that high $F_L$ is essential for the growth of high-quality NiFe. Moreover, we deposited NiFe films on various substrates at room temperature, and the trend of the resistivity depends on the quality of the substrate surface (Fig. 1(d)). As can be shown, NiFe films grown on HF-STO and KTO exhibit the lowest resistivities, suggesting that the substrate surface influences the NiFe film growth.

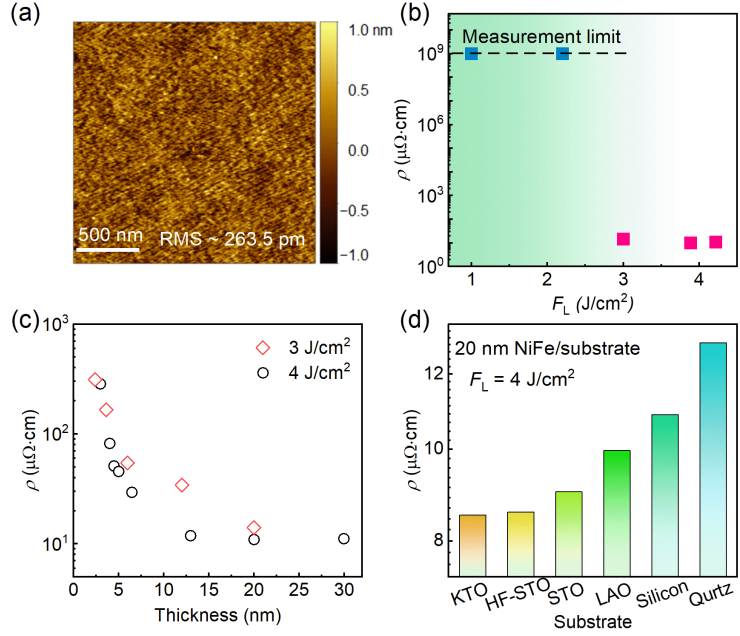

**Figure 1. Growth and characterization of NiFe thin films.** (a) Surface morphology investigations of the NiFe film grown on silicon at a laser fluence ($F_L$) 4 J/cm$^2$ with $t$ = 12 nm (scan scale of 2 × 2 μm$^2$). (b) Room temperature resistivity $\rho$ of 20 nm NiFe films grown on silicon with different $F_L$. (c) Experimental thickness-dependent resistivity of thin NiFe films grown on silicon. (d) Experimental substrate-dependent resistivity of 20 nm thin NiFe films.

The magnetization hysteresis loops (*M-H*) along IP and OOP at room temperature of 20 nm NiFe films grown on silicon at a laser fluence $F_L$ of 4 J/cm$^2$. We observed a sharp magnetic reversal with a strong magnetization (*M*) and small coercivity, as expected in all NiFe thin films except below 6 nm. NiFe films with $t$ < 6 nm show weak hysteresis loops (inset of Fig. 2 (b)), indicating the dead layer of the films. For the $t$ > 6 nm sample, the saturation magnetization $M_s$ is reached in only a few Oe. Figure 2(b) shows the thickness-dependent behaviors of the coercive field ($H_c$) along the IP direction. The $H_c$ decreases from 129.8 Oe to 3.7 Oe in $F_L$ = 3 J/cm$^2$ samples and decreases from 87.5 Oe to 1.8 Oe in $F_L$ = 4 J/cm$^2$ samples (seen in Fig. 2(b)). The IP saturation magnetization ($M_s$) values as a function of the thickness of the NiFe films are shown in Fig. 2(c). For NiFe films grown at $F_L$ = 4 J/cm$^2$, the $M_s$ value was gradually varied from 9.6 emu/cc to 547.5 emu/cc with an increasing $t$ ranging from 3 to 20 nm. For NiFe films grown at $F_L = 3$ J/cm$^2$, the $M_s$ value was gradually varied from 8.9 emu/cc to 270.1 emu/cc with an increasing $t$ ranging from 2.4 to 12 nm. The decrease of $M_s$ with decreasing film thickness is in qualitative agreement with several previous reports on thickness dependence of saturation magnetization in Ni$_{1-x}$Fe$_x$ thin films.[18-21] The magnetic dead layer, which can be deduced by extrapolating $M_s$-$t$ relation to $M_s \rightarrow 0$, is ~2 nm and ~3 nm for $F_L$=3 J/cm$^2$ and 4 J/cm$^2$, respectively. In addition, substrate-dependent $M_s$ of 20 nm thin NiFe films is also observed, as shown in Fig. 2(d). The $M_s$ values vary minorly with different substrates, which indicates the growth is precise. The NiFe films on the LAO substrate exhibit the weakest magnetism with $M_s$ of ~450.5 emu/cc. The difference between each NiFe thin film on various substrates could be due to the different surface roughness. Room temperature $M_s$ values for 20 nm NiFe films have been reported in literature to be about 800 emu/cm$^3$ and 546 emu/cm$^3$ in single and bilayer structures, respectively, deposited by sputtering.[12,18,21,22] The reduced magnetization value may be due to the oxidation and magnetic dead layer effects of the NiFe films, and this fact almost exists in many soft and hard magnetic materials.[23] Our NiFe films exhibit almost similar magnetization with low coercivities compared to those grown by other deposition techniques. The small coercivity and sharp switching behaviour indicate uniaxial anisotropy with an easy axis along the IP direction. Moreover, the NiFe films grown at higher $F_L$ show higher magnetization. Their $M_s$ values are higher and $H_c$ values are lower comparing to NiFe films grown at lower $F_L$ with same $t$.

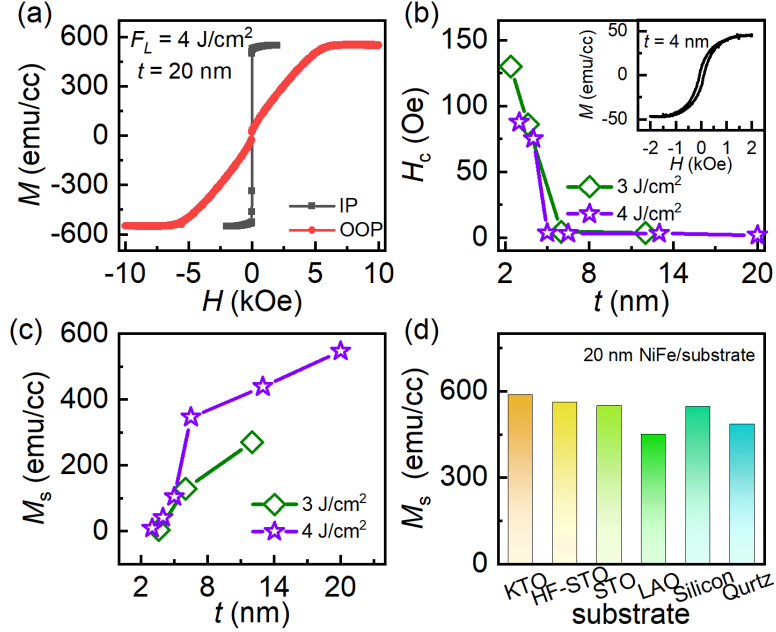

**Figure 2. Magnetic properties of the NiFe films.** (a) In-plane (black square) and out-of-plane (red circle) magnetic hysteresis loops at 300 K for 20 nm NiFe films grown on silicon at a laser fluence of 4 J/cm². Evolution of (b) the coercive field ($H_c$) values and (c) the saturation magnetization ($M_s$) as a function of the NiFe film thickness grown on silicon. The inset in (b) shows the in-plane magnetic hysteresis loop for 4 nm NiFe film. (d) The substrate-dependent saturation magnetization ($M_s$) of thin NiFe films.

To find out the dynamic and damping characteristics at room temperature in the NiFe films, magnetodynamics measurements are performed using the FMR technique. FMR experimental data for NiFe thin film deposited at $F_L$ = 4 J/cm² with $t$ = 20 nm is presented as an example in Figure 3. Figure 3(a) displays the FMR spectra in the wide frequency ($f$) range from 2 to 40 GHz as a function of the external magnetic field ($H$). The circle symbols give the experimental data, while the solid lines show the fits to the derivatives of a Lorentzian function.[24] From the spectra, the resonance field ($H_{res}$) and the linewidth ($\Delta H$) can be extracted by determining the peak-to-peak (inset of Fig. 3(a)).[25] The extracted $\Delta H$ as a function of $f$ is summarized in Figure 3(b). The Gilbert damping could be obtained from the linearly fitted curves (red lines) based on the following equation:

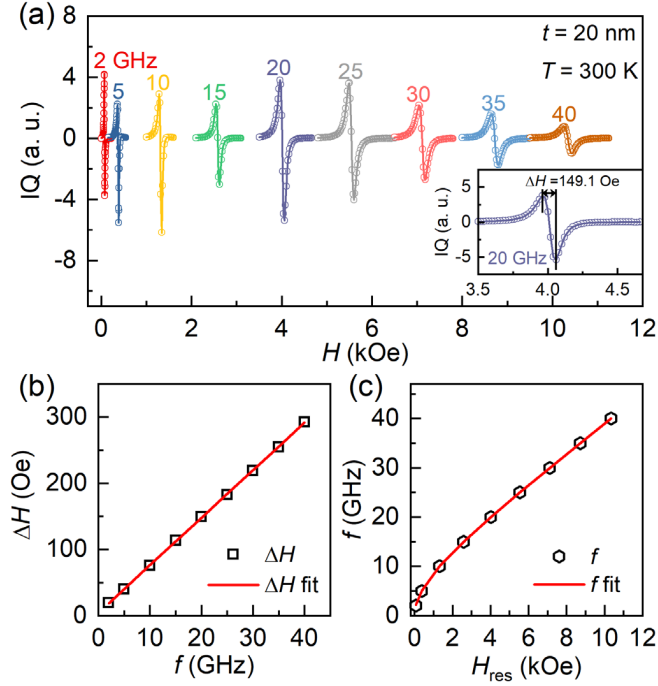

**Figure 3. FMR data of a 20 nm NiFe thin film grow on silicon measured at 300 K with the magnetic field applied in the plane of the NiFe thin film.** (a) The measured FMR signals at different frequencies. The solid lines are the fittings. Inset shows a single FMR spectrum measured at 20 GHz. (b) FMR linewidth as a function of frequency. The effective damping was determined to be 0.01059 from a linear fit based on eq. 1 (red curve). (c) The resonance frequency vs the resonance magnetic field ($H_{res}$) at 300 K. The red line is the fitting result using the Kittel formula (eq. 2).

$$\Delta H = \left(\frac{2\pi}{\gamma}\right)\alpha f + \Delta H_0 \qquad (1)$$

in which $\gamma$ is the geomagnetic ratio, and $\Delta H_0$ is related to the inhomogeneous properties of the NiFe films. The Gilbert damping ($\alpha$) at 300 K is estimated to be 0.01059. We plot the field dependence of the extracted resonance frequency in Fig. 3(c). It is notable that the $H_{res}$ shifts toward higher fields as $f$ increases, which is consistent with the Kittel model,[26]

$$f = (\frac{2\pi}{\gamma})\left[H_{res}(H_{res} + 4\pi M_{eff})\right]^{1/2} \qquad (2)$$

where $M_{eff}$ is the effective magnetization which contains the saturation magnetization and other anisotropy contributions. As the fitted line shown in Fig. 3(c), the $4\pi M_{eff}$ for 20 nm NiFe films is obtained to be ~7.30 kG at 300 K.

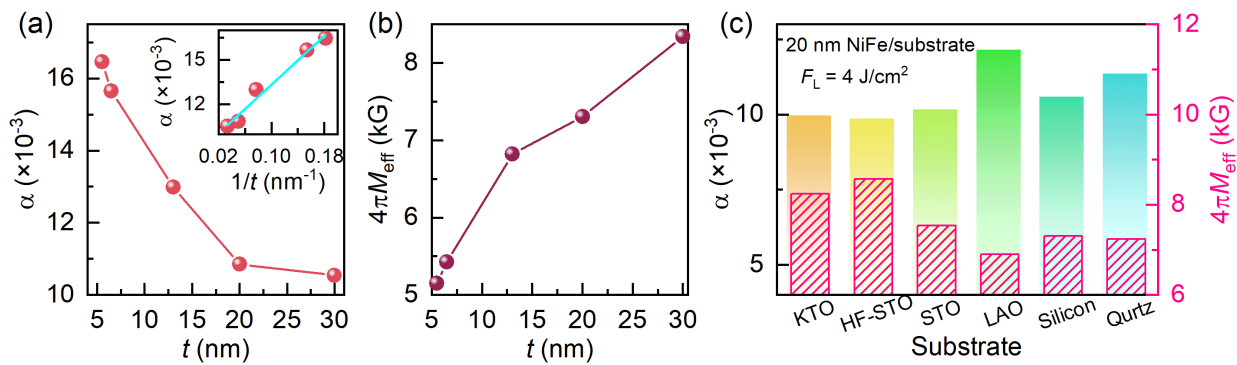

**Figure 4. Gilbert damping and effective magnetization of NiFe thin films.** (a) The Gilbert damping as a function of the NiFe thickness, $t$, measured at 300 K grown on silicon. Inset shows the linear fitting corresponds to eq.3 of Gilbert damping as a function of $1/t$ measured at 300 K. The slope and the intercept are related to the surface contribution and bulk contribution to the total Gilbert damping. (b) Thickness dependence of the effective magnetization $M_{eff}$. (c) Substrate-dependent Gilbert damping (left axis) and effective magnetization (right axis).

The extracted Gilbert damping $\alpha$ as a function of the NiFe thicknesses ($t$) is summarized in Fig. 4(a). As $t$ increases, the Gilbert damping $\alpha$ decreases as expected for magnetic metallic samples, which indicates a surface/interface-enhanced damping for thin NiFe films.[27] As reported, analyzing the damping as a function of $1/t$ can separate the damping contributions due to the bulk and the surface/interface, as shown in the inset of Fig. 4(a).[28] As predicted, it follows the equation of [27, 29, 30]

$$\alpha = \alpha_b + \alpha_s \left(\frac{1}{d}\right) \qquad (3).$$

Here, the $\alpha_b$ and $\alpha_s$ represent the bulk and surface damping, respectively. The best-fitted parameters for $\alpha_b$ and $\alpha_s$ are $0.0099 \pm 0.0004$ and $0.0409 \pm 0.0035$ nm. The $\alpha$ of NiFe film grown at 4 J/cm² is higher than that of at 3 J/cm². For example, it is 0.01494 and 0.01299 for $t =$ 12 nm NiFe film grown at 3 J/cm² and 4 J/cm², respectively. The thickness dependences of the $4\pi M_{eff}$ for NiFe films are shown in Fig. 4(b). The $4\pi M_{eff}$ are obtained to be ~ 5.14 kG to 8.34 kG for *t* from 5.5 nm to 30 nm. In addition, substrate-dependent Gilbert damping $\alpha$ (left axis) and effective magnetization $4\pi M_{eff}$ (right axis) of thin NiFe films are depicted in Fig. 4(c). The NiFe film on LAO presents the largest $\alpha$ and lowest $4\pi M_{eff}$ values.

**Conclusion**

In conclusion, we report the successful growth of high-quality soft ferromagnetic NiFe thin films on oxide/non-oxide substrates at room temperature through PLD employing a KrF pulsed laser source. The fluence of the pulsed laser during deposition affects the magnetization of NiFe thin films. The thin films deposited at high laser fluence exhibits smooth flat surface and excellent ferromagnetic behavior. A few Oe of coercive fields and strong saturation magnetizations were obtained for film thickness above 6 nm. It is found that the magnetization and the damping parameters depend substantially on the film thickness. A noteworthy magnetization value of 547.5 emu/cc is attained at 20 nm of thickness. In addition, results indicate that the ferromagnetic characteristics do depend on the quality of the oxide/non-oxide substrate surface. Our experimental results offer new platforms for research in ferromagnetic thin films and spintronic devices.

**Declaration of Competing Interest**

The authors declare that they have no known competing financial interests or personal relationships that could have appeared to influence the work reported in this paper.

**Acknowledgements**

This research is supported by the Agency for Science, Technology and Research (A*STAR) under its Advanced Manufacturing and Engineering (AME) Individual Research Grant (IRG) (A2083c0054).